\documentclass{aa}

\usepackage{graphicx}
\usepackage{txfonts}
\usepackage{natbib}
\usepackage[usenames, dvipsnames]{color}
\usepackage{array}
\usepackage{xspace}
\usepackage{lscape}
\usepackage{url}
\usepackage{arydshln}
\usepackage{rotating}

\newcommand{\degree}{\ensuremath{^\circ}}

\newcommand{\kms}{\ensuremath{\mathrm{km\ s^{-1}}}\xspace}

\newcommand{\xabs}{{\it xabs}\xspace}

\newcommand{\cie}{{\it cie}\xspace}
\newcommand{\vgau}{{\it vgau}\xspace}

\newcommand{\file}{{\it file}\xspace}
\newcommand{\xmm}{{\it XMM-Newton}\xspace}

\newcommand{\chandra}{{\it Chandra}\xspace}
\newcommand{\nustar}{{\it NuSTAR}\xspace}

\newcommand{\hst}{{HST}\xspace}

\begin{document}

\title{X-ray narrow emission lines from the nuclear region of NGC 1365}
\subtitle{}

\author{M. Whewell \inst{1} 
\and G. Branduardi-Raymont \inst{1}
\and M. J. Page \inst{1}
}

\institute{Mullard Space Science Laboratory, University College London, Holmbury St. Mary, Dorking, Surrey, RH5 6NT, UK \\{\email{m.whewell@ucl.ac.uk}} 
}

\date{Received date /
Accepted date}

\abstract{NGC 1365 is a Seyfert 2 galaxy with a starburst ring in its nuclear region. In this work we look at the \xmm Reflection Grating Spectrometer (RGS) data from four 2012-13, three 2007 and two 2004 observations of NGC 1365, in order to analyse and characterise in a uniform way the soft X-ray narrow-line emitting gas in the nucleus.}{We characterise the narrow-line emitting gas visible by \xmm RGS and make comparisons between the 2012-13 spectra and those from 2004-07, already published.}{This source is usually absorbed within the soft X-ray band, with a typical neutral column density of $>$1.5 x 10$^{23}$ cm$^{-2}$, and only one observation of the nine we investigate shows low enough absorption for the continuum to emerge in the soft X-rays. We stack all observations from 2004-07, and separately three of the four observations from 2012-13, analysing the less absorbed observation separately. We first model the spectra using gaussian profiles representing the narrow line emission. We fit physically motivated models to the 2012-13 stacked spectra, with collisionally ionised components representing the starburst emission and photoionised line emission models representing the AGN line emission. The collisional and photoionised emission line models are fitted together (rather than holding either one constant), on top of a physical continuum and absorption model.}{The X-ray narrow emission line spectrum of NGC 1365 is well represented by a combination of two collisionally ionised (kT of 220$\pm10$ and 570$\pm15$\,eV) and three photoionised ($\log \xi$ of 1.5$\pm0.2$, 2.5$\pm0.2$, 1.1$\pm0.2$) phases of emitting gas, all with higher than solar nitrogen abundances. This physical model was fitted to the 2012-13 stacked spectrum, and yet also fits well to the 2004-07 stacked spectrum, without changing any characteristics of the emitting gas phases. Our 2004-07 results are consistent with previous emission line work using these data, with five additional emission lines detected in both this and the 2012-13 stacked spectra. We also estimate the distance of the X-ray line-emitting photoionised gas from the central source to be $<$300\,pc.}{}

\authorrunning{M. Whewell et al.}
\titlerunning{NGC 1365 X-ray lines}
\maketitle

\section{NGC 1365 - Introduction}
\label{NGC1365_intro}

\begin{table*}
\begin{minipage}[t]{\hsize}
\setlength{\extrarowheight}{3pt}
\caption{\xmm and \chandra observations of the nuclear region of NGC 1365 and where the data have been analysed. The final four \xmm observations were taken simultaneously with \nustar.}
\label{obs_table}
\centering
\renewcommand{\footnoterule}{}
\begin{tabular}{l l c c | c}
\hline \hline
Telescope & Start of obs. & Obs. ID & {Duration (ks) \footnote{For \chandra observations the scheduled length is quoted}} & References \\
\hline
\chandra & 2002-12-24 & 3554 & 15 & 1, 9 \\
\xmm & 2004-01-17 & 0205590301 & 59.7 & 2, 7, 5, 10 \\
\xmm & 2004-07-24 & 0205590401 & 68.8 & 2, 5, 10 \\
\chandra & 2006-04-17 & 6868 & 15 & 3, 4 \\
\chandra & 2006-04-20 & 6869 & 15 & 3, 4 \\
\chandra & 2006-04-23 & 6870 & 15 & 3, 4 \\
\chandra & 2006-04-10 & 6871 & 15 & 3, 4 \\
\chandra & 2006-04-12 & 6872 & 15 & 3, 4 \\
\chandra & 2006-04-12 & 6873 & 15 & 3, 4 \\
\xmm & 2007-06-30 & 0505140201 & 128.9 & 6, 5, 10 \\
\xmm & 2007-07-02 & 0505140401 & 131.1 & 6, 5, 10 \\
\xmm & 2007-07-04 & 0505140501 & 130.9 & 6, 5, 10 \\
\chandra & 2012-04-09 & 13920 & 90 & 11, 13 \\
\chandra & 2012-04-12 & 13920 & 120 & 11, 13 \\
\xmm & 2012-07-25 & 0692840201 & 138.5 & 8, 10, 12, 14 \\
\xmm & 2012-12-24 & 0692840301 & 126.2 & 10, 12, 14 \\
\xmm & 2013-01-23 & 0692840401 & 133.6 & 11, 10, 12, 14 \\
\xmm & 2013-02-12 & 0692840501 & 134.7 & 10, 12, 14 \\
\hline
\hline
\end{tabular}
\tablebib{
(1)~\citet{Risaliti:2005dj}; (2) \citet{Risaliti:2005kd}; (3) \citet{2007ApJ...659L.111R}; (4) \citet{Wang:2009jh}; (5) \citet{Guainazzi:2009fv}; (6) \citet{Risaliti:2009ic}; (7) \citet{Risaliti:2009cx}; (8) \citet{2013Natur.494..449R}; (9) \citet{Connolly:2014ft}; (10) \citet{2014MNRAS.441.1817P}; (11) \citet{Braito:2014ct}; (12) \citet{Walton:2014fc}; (13) \citet{2015MNRAS.453.2558N}; (14) \citet{Rivers:2015fy}
}
\end{minipage}
\end{table*}

\object{NGC 1365} (z $=$ 0.0055) has an interesting observational classification history. Optically it has been called a Seyfert 1.5, 1.8 or 2 by different authors \citep[e.g.][respectively]{1980A&A....87..245V,1995ApJ...454...95M,1993ApJ...418..653T}. Most recent X-ray papers have classified NGC 1365 as a Seyfert 2 due to the column density of neutral material \citep[$>1.5 \times 10^{23}$ cm$^{-2}$,][]{2007ApJ...659L.111R,Risaliti:2009cx} covering its X-ray emission \citep[intrinsic L$_{2-10\,keV} \sim10^{42}$\,erg\,s$^{-1}$][]{Risaliti:2005dj}.

Notably, especially for this work, there is a nuclear starburst with a diameter of 10'', which is resolved in optical wavelengths into multiple compact star clusters \citep[using the \hst Faint Object Camera, ][]{1997A&A...328..483K}. This is often referred to as a `ring', and we continue to do so in this work, although there is some evidence from radio emission that the ring is incomplete \citep{Beck:2005cb}.
Starburst regions are known to produce soft X-rays primarily from collisionally excited, shock-heated gas, and hard X-rays mainly from accreting compact objects such as X-ray binaries. Any interpretation of soft X-ray emission from AGN photoionised gas in NGC 1365 must take this surrounding starburst emission into account.

It is well established that the variability seen in the spectrum of NGC 1365 is caused by variations in photoelectric absorption. Recently, \cite{2014MNRAS.441.1817P} showed this in a model independent way by using Principal Component Analysis (PCA). Comparing PCA results of analysis of NGC 1365 to results from model spectra, varied in different physical ways, showed that variable absorption causes  $>\,90\,\%$ of the variability in NGC 1365's observed spectra.
The details of the complex phases of X-ray absorbing gas found in this source are summarised below.


\textit{Emission line gas.} Gas photoionised by the central AGN of NGC 1365 has been studied in both optical and X-ray bands. Optical line ratios in an asymmetrical conical region suggests photoionisation by the AGN \citep{Veilleux:2003fe,1999A&A...346..764S}.

NGC 1365 was the first AGN where high-quality, high-resolution soft X-ray data showed emission lines dominated by collisionally ionised gas \citep{Guainazzi:2009fv}. In their analysis of a $\sim$500\,ks stacked spectrum collected by RGS between 2004 and 2007, \cite{Guainazzi:2009fv} model the emission lines with two collisional components (kT$\sim$300 and $\sim$640\,eV) and one loosely constrained photoionised component ($N_H \geq 10^{22}$\,cm$^{-2}$, $\log$\,U\,$= 1.6^{+0.3}_{-0.4}$, $n_e \leq 10^{10}$\,cm$^{-3}$).

Studies since \cite{Guainazzi:2009fv} have confirmed the two collisional phases of emission line gas.
Characterising the photoionised emission in the presence of this collisional emission has been difficult, but restricting the spatial dimension of \chandra HETGS spectra allowed \cite{2015MNRAS.453.2558N} to reduce the contribution of collisional emission from the circumnuclear starburst ring in their spectra compared to previous analyses \citep[e.g.][]{Guainazzi:2009fv}. They confirm the presence of photoionised emission, as well as two collisionally ionised phases with different temperatures to those found by \cite{Guainazzi:2009fv} ($\sim$150\,eV and $\sim$1200\,eV as opposed to $\sim$300\,eV and $\sim$640\,eV). \cite{2015MNRAS.453.2558N} accept that the spectral quality of their data is limited by lack of photon counts and therefore a detailed characterisation of the photoionised emission lines cannot be done with this dataset. The authors loosely contrain the temperature and density of the emitting gas to $T < 3 \times 10^6$\,K and $n < 10^{13}-10^{14}$\,cm$^{-3}$ respectively. They could fit the spectrum with three photoionised emission models at ionisations of $\log \xi > 4.1, \sim3.1\pm0.3$ and $< 0.6$ but recommend interpretation of this as simply proof of the existence of a large range of ionisation states. 

Using the \chandra data, \cite{2015MNRAS.453.2558N} estimated the location of the photoionised gas to be `across the virtual BLR/NLR boundary', based on some evidence of broadening of the emission lines ($\sim1000\,$\kms). \cite{Guainazzi:2009fv} tentatively suggested the inner face of the photoionised gas could be $\geq 0.75$\,pc from the source, potentially placing it within the BLR. \cite{Braito:2014ct} use an \xmm observation in 2013 (RGS and EPIC) and also conclude that some emission line gas is located within the BLR, as they see the Mg\,XII\,Ly-$\alpha$ emission line increase in strength as their measured absorption decreases. They interpret this as either the uncovering of emission from within their most variable absorber ($r < 10^{17}$\,cm) or ionised gas responding to a change in its illuminating continuum, which (given the likely timescales) would have to be a compact region and then would again be likely located at $r \sim 10^{15}$\,cm from the source.

Finally, \cite{Braito:2014ct} suggest that the Mg\,XII emission and absorption lines in the highest flux part of their observation are reminiscent of a classical P-Cygni profile, and therefore could hint at a common origin for both from an outflowing wind.


\textit{Multiple layer X-ray absorbers.} The X-ray absorption complex within NGC 1365's nuclear region has been studied in detail. In addition to the original neutral absorber characterised by \cite{Risaliti:2005dj}, shown to cause the main variability seen from the source \citep[][]{Risaliti:2005dj,Risaliti:2009ic,2014MNRAS.441.1817P}, many other absorption layers have also been detected with X-ray data.

\cite{Risaliti:2005kd} found absorption lines between 6.7 and 8.3\,keV with \xmm EPIC data of the source in a Compton-thin state, which they attribute to an absorber originating from an accretion disk wind.

By analysing one 120\,ks (2013) \xmm observation in detail, \cite{Braito:2014ct} find at least two ionised absorber phases (and evidence for a third), which are needed to fit the RGS spectrum, as well as a neutral absorber needed to fit the EPIC spectrum (not statistically required in the RGS band). The lowest ionisation phase of these three is found to be partially covering and variable in column density over timescales as short as 40\,ks (within the 120\,ks observation).
The three ionised absorber phases they use are each responsible for different signatures within the data. Their highest ionisation phase ($\log \xi = 3.8$), which causes the absorption features around the Fe\,K line complex \citep[the absorption lines found by][]{Risaliti:2005kd}, is outflowing by 3900\,\kms, and is only needed by the EPIC spectrum. Their mid ionisation phase ($\log \xi = 2.1$) produces Mg, Si and S absorption lines, fitted in both RGS and EPIC spectra, and with an RGS measured outflow velocity of 1200\,\kms. Their lowest ionisation phase ($\log \xi < 1$) causes spectral curvature in the EPIC spectrum and the Fe UTA in the RGS spectrum.
\cite{Braito:2014ct} propose that the three ionised absorber phases are part of a disk wind, and the lowest ionisation phase could be identified as the partially covering neutral absorber previously detected by \cite{Risaliti:2005dj}, located around the distance of the BLR \cite[r\,$<\,10^{16}$\,cm; e.g.][]{Brenneman:2013kw,Risaliti:2005kd}.

The same four 2012-13 joint \xmm and \nustar observations analysed by \cite{Walton:2014fc}, \cite{2013Natur.494..449R} and \cite{Braito:2014ct} were also studied by \cite{Rivers:2015fy}, to untangle any extra complexities and layers of neutral absorption. The \cite{Rivers:2015fy} model also includes absorption lines to represent the ionised absorption discovered around 6-8\,keV by \cite{Risaliti:2005kd}.
They observed that when the source becomes uncovered by its high column density partially covering absorber (the covering fraction decreases), such as in observations 2, 3 and 4 of this set (in Dec 2012, Jan 2013 and Feb 2013 respectively), an additional layer of fully covering neutral absorption is needed to fit the spectrum.

Their model for a partially covering layer of absorption varies substantially within and between observations 1, 3 and 4. Its behaviour during observations 1 and 4 is similar, with column density changes but no covering fraction changes, and here \cite{Rivers:2015fy} call it a ``partial-covering `high column density' absorber''. Its behaviour in observation 3 is very different, with a rapid drop of both covering factor and column density, which \cite{Rivers:2015fy} refer to as an uncovering of the source, and call this a ``patchy partial-covering'' absorber instead. They argue that these two different behaviours are evidence that the one partial covering model they use actually tracks different layers of the absorption, in reality distinct from each other.

In this work we look at the RGS data from all four of these 2012-13 \xmm observations, and at the \xmm observations from 2004 and 2007, in order to analyse in a uniform way and further characterise the X-ray narrow-line emitting photoionised gas. This paper is structured as follows: in Sect. \ref{NGC1365_obsdata} we describe the observations and data reduction process; in Sect. \ref{NGC1365_mywork} we go on to present an analysis of the emission line spectrum from NGC 1365, firstly using gaussian line models (Sect. \ref{gaussian_fit}) before moving on to a combination of physically motivated emission models (Sect. \ref{physicalmodels_sect}). We discuss the implications of our findings in Sect. \ref{NGC1365_discussion}, and finally conclude in Sect. \ref{NGC1365_conclusion}.

\section{Observations and Data Reduction}
\label{NGC1365_obsdata}

In this paper we consider the nine \xmm observations of NGC 1365 we obtained from the \xmm Science Archive; two from 2004, three from 2007, two from 2012 and two from 2013 (see Table \ref{obs_table}). We use only the RGS data from each observation as they are the only data of high enough resolution to study the narrow emission lines.

Using the standard data reduction pipeline, SAS, it is possible to combine spectra from the same or both RGS chains (RGS1 and/or RGS2), over many observations, as long as they are all the same spectral order (1st or 2nd). It is not possible to combine all data from both detectors and both spectral orders over many observations.
In order to do this, all the \xmm RGS data used in this paper have therefore been reduced at a more advanced level than the standard SAS pipeline.
This is a variation on the RGS data reduction process described in \cite{Kaastra:2011bo},
which also allows combinations of spectra with different occurrences of bad pixels and from the multi-pointing mode \citep[such as for the NGC 5548 campaign, described in][]{Mehdipour:2014ip}.

While the data reduction process within \cite{Kaastra:2011bo} allows both spectral orders from RGS observations to be combined, the authors state that only $\chi^2$ statistics may be used to analyse the spectra produced, as information on the background level is not carried through the process. This is because fluxed RGS spectra created by SAS do not include background information for each bin.
Cash statistics (C-statistic) cannot be used correctly without this information and therefore is not appropriate for use on any spectra produced this way.

The individual spectra of NGC 1365 used in this work have low numbers of counts per bin (the average number of source and background counts per bin can be as low as 5), and therefore $\chi^2$ is not the appropriate goodness-of-fit parameter. In order to use C-statistics the data reduction process had to be altered from that described in \cite{Kaastra:2011bo}, so the background and source count information is retained.

The Observation Data Files (ODFs) for each observation are processed with the SAS metatask \textit{rgsproc} (using SASv14) to produce background-subtracted spectra and response matrices for each spectral order (1st and 2nd) and each chain (RGS1 and RGS2). This therefore gives four source spectra per observation. The background-subtracted spectra still include information on the background level for each bin.
High-background time intervals were filtered out by a threshold of 0.15\,counts\,s$^{-1}$ on the background lightcurve extracted from a background region of CCD-9, the CCD closest to the optical axis and therefore most affected by background flares.

The spectra from the SAS are then converted into {\sc spex} \citep{1996uxsa.conf..411K}
format using the auxiliary program \textit{trafo}, then the first order spectra from both detectors are fitted together with a preliminary model using {\sc spex}. This model only needs to empirically describe the data, it does not need to be physically motivated. 

At this point we use the empirical model to convert all four spectra for that observation into fluxed spectra (in photons\,s$^{-1}$\,m$^{-2}$\,\AA$^{-1}$, instead of the previous counts\,s$^{-1}$\,m$^{-2}$\,\AA$^{-1}$, and including information on background). 
Then they can be combined, taking into account the detector and spectral order combinations. The second order spectra are weighted differently to the first orders, to reflect the greater signal to noise and number of counts in the first order spectra.

Now we have one spectrum for each observation, which can be stacked using a process similar to the {\it rgsfmat} auxilary function in {\sc spex}, which additionally takes into account the length of each individual observation. This also creates the relevant response files for both the stacked and individual observations.

The RGS data are analysed in the range 11-38\,\AA, with a bin size of 0.033\,\AA, which over-samples the RGS resolution element of $\sim$0.07\,\AA. All the spectra in this paper are shown as background-subtracted and in the observed frame. The spectral analysis and modelling for this work was done with {\sc spex} (v.2.06.01).

\section{Analysis}
\label{NGC1365_mywork}

We compared the 2012-13 observations (1, 2, 3 and 4) by eye, and determined that observations 1, 2 and 4 are very similar above 11\,\AA. We stacked these three observations together to begin our analysis, with a combined exposure time of $\sim$355\,ks. Observation 3 has a lower level of absorption than the others \citep{Braito:2014ct}. 

We also reduced and stacked the five observations taken in 2004 and 2007, giving a stacked spectrum of $\sim$419\,ks, in order to compare our results with \cite{Guainazzi:2009fv} and to check any variation of the emission lines over time. We return to the individual observations to look for variations between their emission lines after we have established a baseline model (Sect. \ref{ind_obs}). 

When using C-statistic to fit models in {\sc SPEX}, as well as the calculated C-statistic value, {\sc SPEX} also gives the expected C-statistic value and its r.m.s. uncertainty to help the user determine goodness-of-fit (see the {\sc SPEX} manual for futher information).
For the 2012-13 stacked spectrum the expected C-statistic is $911\pm43$ and for the 2004-07 stacked spectrum it is $907\pm43$.
The high spectral resolution of RGS ($\lambda / \Delta\lambda\sim340$ at $22\,\AA$; \citealp{denHerder:2001fv}) compared to lower resolution X-ray spectroscopy leads to more detailed models being needed to fully describe the data. Our knowledge of many aspects of these models is not yet complete (for example the possible presence of additional astrophysical components in the line of sight towards the source and imperfect atomic data) and we also have to deal with imperfect instrumental calibration, systematic effects and statistical fluctuations, so there will necessarily be differences between the data and the best physical model that can be found.
As discussed in \cite{Blustin:2002hj}, the use of goodness-of-fit statistics when modelling RGS data must be coupled with careful consideration of the physical meaning of the models, as physical self-consistency is more important than reaching a nominally ``acceptable'' goodness-of-fit value.

\subsection{Testing the 2012-13 stacked spectrum with continuum model and gaussian lines}
\label{gaussian_fit}

As \cite{Braito:2014ct} have previously analysed the Jan 2013 observation (Obs. 3, the least absorbed of the four 2012-13 observations), we began our analysis using their continuum and absorption model. Their use of EPIC data allows much better contraints of the continuum than is possible using just the RGS data. As observation 3 has lower absorption than the other 2012-13 observations, it reveals ionised absorber phases not previously seen.
For this we used a power law model, absorbed by three ionised absorbers ($\log \xi$ of 3.77, 2.15 and 0.17, as in their Table 4). We allowed our power law normalisation to vary, and fixed $\Gamma$ to a value of 2.1 to match their EPIC fits. At this stage we have not fitted any of the emission lines, so have an unacceptable C-statistic value of $>$7000.

\begin{table*}
\begin{minipage}[t]{\hsize}
\setlength{\extrarowheight}{0.3pt}
\caption{Narrow emission line best fit intensities from the 2012-2013 stacked spectrum and 2004-2007 stacked spectrum. A star marks uncertainty in the assigned transition.  Overall C-statistic $=$ 1445 (2012-13) and 1548 (2004-07)}
\label{NLgaussian_table}
\centering
\renewcommand{\footnoterule}{}
\begin{tabular}{l c | c c c}
\hline \hline
{Line Transition \footnote{r, i and f represent the resonance, intercombination and forbidden lines of an He-like triplet, respectively}} & Rest $\lambda$ & 2012-13 Intensity & 2004-07 Intensity & 2004-07 (G09) Intensity \\ 
& (\AA) & (10$^{-5}$ ph cm$^{-2}$ s$^{-1}$) & (10$^{-5}$ ph cm$^{-2}$ s$^{-1}$) & (10$^{-5}$ ph cm$^{-2}$ s$^{-1}$) \\ 
\hline
Ne X 1s-2p & 12.14 & 	0.85$\pm$0.11 & 1.12$\pm$0.11 & 1.0$\pm$0.2 \\ 
Fe XXI * & 12.29	 & 	0.20$\pm$0.09 & 0.21$\pm$0.08 & - \\ 
Ne IX r & 13.45 & 		0.99$\pm$0.13 & 1.28$\pm$0.11 & 0.42$\pm$0.18 \\ 
Ne IX i & 13.55 & 		$<$0.13 & $<$0.08 & $<$0.33 \\ 
Ne IX f & 13.70 & 		0.90$\pm$0.10 & 1.02$^{+0.08}_{-0.02}$ & 0.40$\pm$0.12 \\ 
O VIII RRC & 14.23 &  	0.44$\pm$0.07 & 0.59$\pm$0.07 & 0.42$\pm$0.13 \\ 
Fe XVIII * & 14.41 &  	0.33$\pm$0.10 & 0.20$\pm$0.05 & - \\ 
Fe XVII 3d-2p & 15.03 & 	1.63$\pm$0.09 & 1.66$\pm$0.09 & 1.49$\pm$0.17 \\ 
O VIII 1s-4p / & 15.18 	& & & \\
Fe XVII 3d-2p & 15.26 & 	1.18$\pm$0.10 & 1.02$^{+0.08}_{-0.03}$ & - \\ 
O VIII 1s-3p & 16.01 & 	0.87$\pm$0.08 & 0.81$\pm$0.07 & 0.75$\pm$0.13 \\ 
O VII RRC / & 16.77 	& & & \\
Fe XVII 2s-2p & 16.78 &  1.05$\pm$0.08 & 1.12$\pm$0.08 & 0.47$\pm$0.15 \\ 
Fe XVII 3s-2p & 17.07 & 	2.31$\pm$0.12 & 2.73$\pm$0.11 & 1.89$\pm$0.19 \\ 
N VII RRC / & 18.59 	& & & \\
O VII 1s-3p & 18.63 & 	0.25$\pm$0.07 & 0.51$\pm$0.10 & - \\ 
O VIII 1s-2p & 18.97 & 	2.65$\pm$0.13 & 2.97$^{+0.12}_{-0.05}$ & 2.5$\pm$0.2 \\ 
O VII r & 21.60 & 		1.23$\pm$0.16 & 1.01$\pm$0.15 & 1.0$\pm$0.2 \\ 
O VII i & 21.81 & 		0.50$\pm$0.13 & 0.58$\pm$0.13 & 0.6$\pm$0.2 \\ 
O VII f & 22.10 & 		2.04$\pm$0.19 & 2.23$\pm$0.19 & 2.4$\pm$0.3 \\ 
N VII 1s-2p & 24.78  & 	1.35$\pm$0.13 & 1.60$\pm$0.12 & 1.2$\pm$0.2 \\ 
N VI 1s-3p & 24.90 & 	0.66$\pm$0.10 & 0.47$\pm$0.08 & - \\ 
C VI 1s-4p & 26.99 & 	0.27$\pm$0.09 & 0.12$^{+0.10}_{-0.06}$ & 0.40$\pm$0.18 \\ 
C VI 1s-3p & 28.44 & 	0.23$\pm$0.13 & 0.61$\pm$0.13 & 0.6$\pm$0.3 \\ 
N VI r & 28.79 & 		0.34$\pm$0.18 & 0.26$\pm$0.14 & 0.7$\pm$0.4 \\ 
N VI i & 29.08 & 		0.24$\pm$0.15 & 0.25$\pm$0.14 & $<$0.6 \\ 
N VI f & 29.53 & 		1.70$\pm$0.20 & 1.61$\pm$0.18 & 1.8$\pm$0.3 \\ 
C VI 1s-2p & 33.74 & 	1.30$\pm$0.24 & 1.86$^{+0.13}_{-0.18}$ & 1.4$\pm$0.5 \\ 
\hline
\hline
\end{tabular}
\end{minipage}
\end{table*}

\begin{figure*}[!]
\centering
\resizebox{\hsize}{!}{\includegraphics[angle=0]{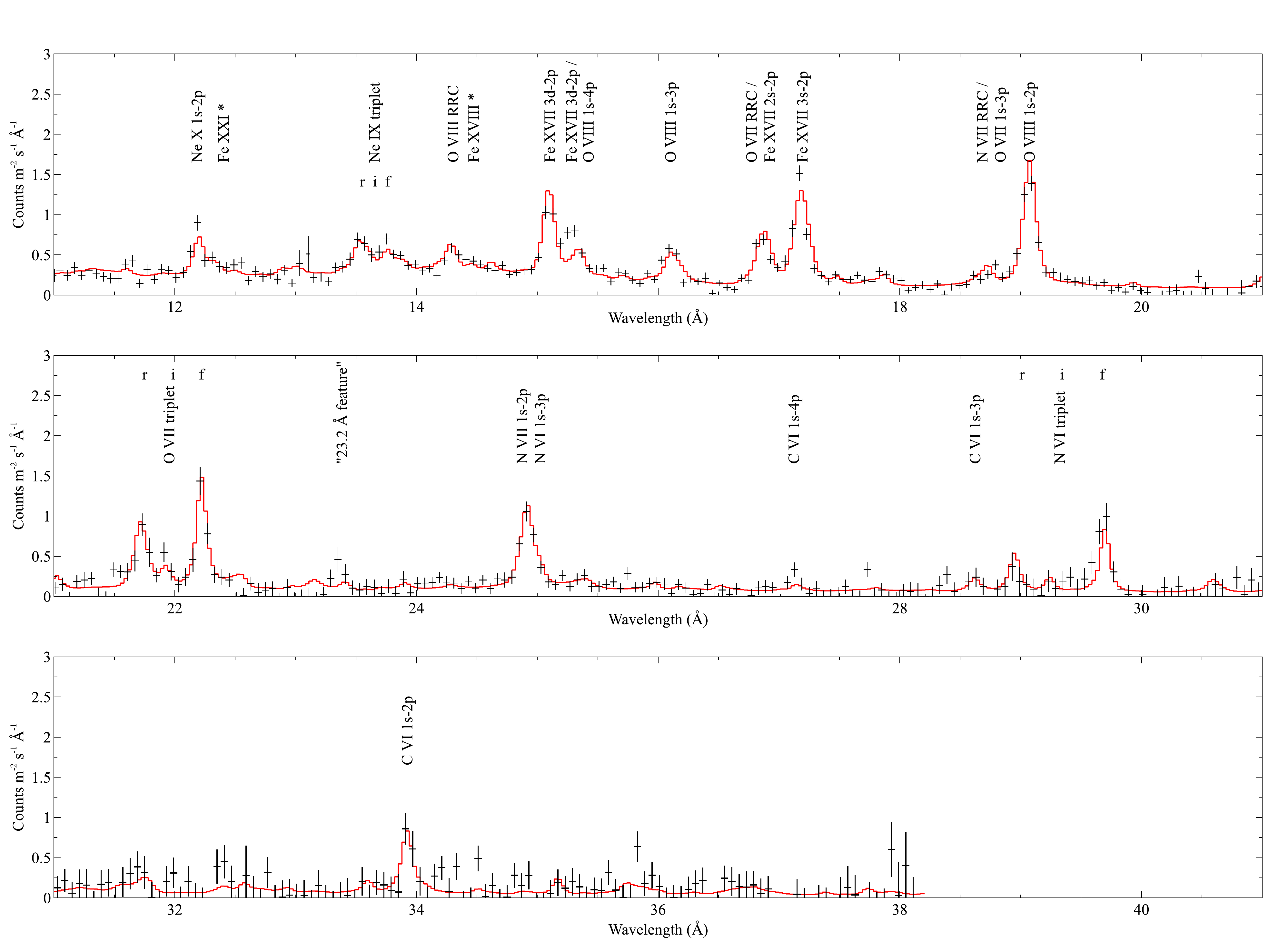}}
\caption{2012-13 RGS spectrum and line labels for detected emission lines. r, i and f represent the resonance, intercombination and forbidden lines of a He-like triplet, respectively. A * indicates uncertainty in the line identification. The data are binned by two for display purposes. The data are shown in black, and the model in red. For details of the model shown here, see Sect. \ref{physicalmodels_sect}.}
\label{NGC1365_dataandlines}
\end{figure*}

We then fitted gaussian lines to the data, adding a gaussian if the best fit C-statistic improves by 6.25, corresponding to a 90\% confidence level for three extra free parameters (normalisation, width and central wavelength of each gaussian model). These gaussian models are only absorbed by our Galactic N$_H$, and none of the absorption intrinsic to NGC 1365. For the He-like triplets in this wavelength range we included three components (resonance, intercombination and forbidden) if at least one was detected at the 90\% level. These `triplets' really include four components, but at RGS resolution the two components of the intercombination line cannot be separated. We also included any relevant lower order transitions for an ion where we detect a higher order transition at the 90\% level. For example, we include C VI 1s-3p (28.44\,\AA), although its detection did not fulfil this criterion, because we detect C VI 1s-4p (26.99\,\AA) at the stated level.

We explored whether the parameters of the power law ($\Gamma$) and absorber components ($\xi$ and N$_{H}$) change when they are fitted, rather than directly using \cite{Braito:2014ct}'s values. We expect changes to the absorption, because there are visible differences between observation 3 (which \cite{Braito:2014ct} analysed in detail) and observations 1, 2 and 4, which we have stacked here.
We fixed the gaussian lines to their current best fit values and first allowed the $\Gamma$ of the power law to vary. We found this unconstrained by the RGS data, as expected, so fixed it back to 2.1. We then also allowed the ionisations or the column densities of the absorbing components (as \xabs models) to vary. Finally we allowed both the ionisations and column densities of the \xabs models to vary at the same time.

The highest ionisation WA component is constrained by \cite{Braito:2014ct} using EPIC data, so it is unsurprising that this becomes very loosely constrained using only RGS data. The low and mid ionisation components increase in ionisation (from $\log \xi$ 0.17$\pm$0.1 to 0.70$\pm$0.1 and 2.15\,$\pm$\,0.1 to 3.05\,$\pm$\,0.4 respectively) and also change in column density (from 1.1$\pm$0.1\,$\times 10^{22}$\,cm$^{-2}$ to 2.7$\pm$0.3\,$\times 10^{22}$\,cm$^{-2}$ and 1.1$\pm$0.4\,$\times 10^{22}$\,cm$^{-2}$ to $<$0.5\,$\times 10^{22}$\,cm$^{-2}$).
We also find an upper limit to the presence of neutral absorption, which has a column density of $<$0.023\,$\times 10^{22}$\,cm$^{-2}$, and the inclusion of this is what renders the mid-ionisation WA phase statistically unnecessary. This is unsurprising as \cite{Braito:2014ct} analysed the least absorbed of the four 2012-13 observations, and the lower neutral absorption allowed them to detect the previously unseen $\log\,\xi \sim 2$ absorbing phase.

So we fixed the highest ionisation absorber's ionisation and column density, as well as the mid ionisation absorber's ionisation value to the parameters reported by \cite{Braito:2014ct} and allowed the other ionised absorber parameters, the neutral absorber column density, as well as the power law normalisation free to vary. These parameters remained consistent with those in the previous paragraph, with a C-statistic of 1445, an improvement on the previous C-statistic (including gaussian emission lines) of $>$ 150, with only four additional free parameters.

We took this overall model for the 2012-2013 observations and refitted both the continuum, absorption and emission lines to a stacked spectrum combining the two 2004 and three 2007 observations. In this case we fixed the power law $\Gamma$ value to 2.6 to better represent the intrinsic continuum at that time, as we know from \cite{Risaliti:2005kd} and \cite{Risaliti:2009ic} that the powerlaw $\Gamma$ was steeper in the 2004 and 2007 observations ($\sim2.6$, compared to $2.1$ during the 2012-13 observations). We allowed the same absorption parameters to vary as described above.
These are our best fit models, and the emission line intensities are given in Table \ref{NLgaussian_table}, where we also compare our values to \cite{Guainazzi:2009fv}'s results. This comparison is discussed in Section \ref{NGC1365_compareotherwork}. The line identifications can be seen with the data in Fig. \ref{NGC1365_dataandlines}, and all our detections are consistent with being at rest relative to the host galaxy within 3$\sigma$ uncertainties.

\subsection{Fitting individual observations}
\label{ind_obs}

We took the best fit models to the stacked spectra described in the above section and fitted them to the relevant individual observations.
First we refitted the continuum and absorption parameters (with the emission lines fixed) and then the emission lines, continuum and absorption parameters together.

The absorption varies between observations, as expected \citep[e.g.][]{2014MNRAS.441.1817P}. As absorption variability in this source is a well-studied and complicated phenomenon, we do not comment on it further in this paper. We have only briefly modelled it here to ensure we have an appropriate continuum and absorption model underlying our emission model, and have visually inspected each observation's model to ensure the continuum and absorption are well represented.

All individual observations are fitted well, and any potential variations of the emission lines are smaller than their 3$\sigma$ uncertanties within the two epochs we stacked (2004-07 and 2012-13).

\subsection{Emission Line Comparison}
\label{NGC1365_compareotherwork}

We compared the emission lines found in our 2012-13 (observations 1, 2 and 4) and our 2004-07 stacked spectra with the emission lines found by \cite{Guainazzi:2009fv} in their 2004-07 stacked spectrum (see Table \ref{NLgaussian_table}).

We find emission lines at all wavelengths where \cite{Guainazzi:2009fv} report emission lines.
We also find evidence for five additional emission lines.

\begin{figure*}[!]
\centering
\resizebox{\hsize}{!}{\includegraphics[angle=0]{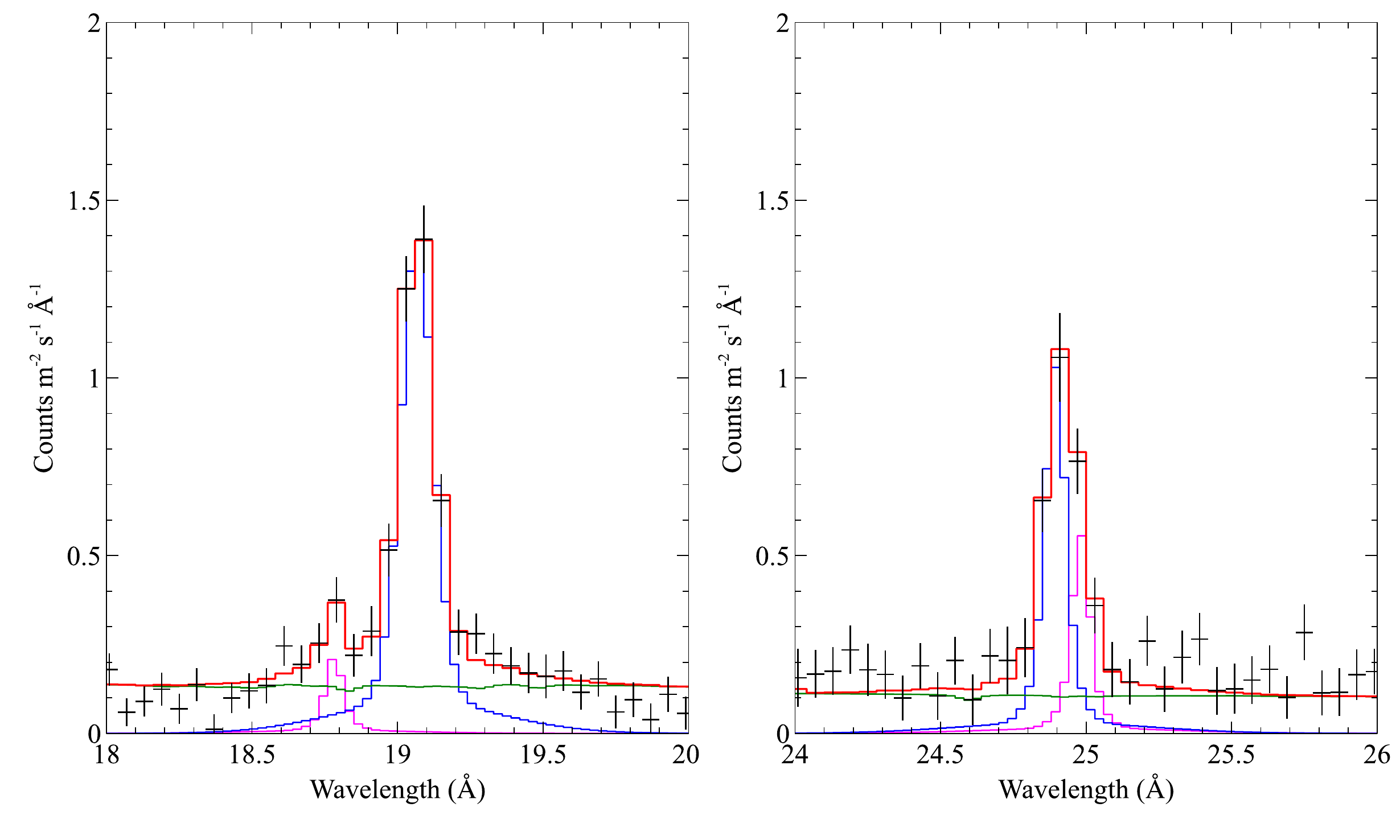}}
\caption{Location of the two new 1s-3p emission lines found, with gaussian components shown. In both panels the data are shown in black, the total model in red and the absorbed powerlaw component in green. Left: Gaussian components for O VIII 1s-2p and O VII 1s-3p shown in blue and pink respectively. Right: Gaussian components for N VII 1s-2p and N VI 1s-3p shown in blue and pink respectively.}
\label{NGC1365_2newlines}
\end{figure*}

We have labelled the additional emission lines found at 18.66$^{+0.02}_{-0.08}$\,\AA\,and 24.85$\pm$0.02\,\AA\,as O VII 1s-3p (rest wavelength: 18.63\,\AA) and N VI 1s-3p (rest wavelength: 24.90\,\AA) lines respectively. Both these lines suffer from blending with nearby, stronger emission lines (O VIII 1s-2p and N VII 1s-2p respectively), as shown in Fig. \ref{NGC1365_2newlines}. There could also potentially be a nitrogen RRC contribution at 18.59\,\AA\, contaminating the O VII 1s-3p line, and this is difficult to disentangle.

The emission line at 15.22$\pm$0.01\,\AA\, could be identified with either O VIII Ly-$\gamma$ (rest wavelength: 15.176\,\AA) or Fe XVII 3d-2p (rest wavelength: 15.262\,\AA), and we consider this likely to be a blend of both. If we assume this line is O VIII Ly-$\gamma$ then we measure a stronger intensity for this than for O VIII Ly-$\beta$ (rest wavelength: 16.01\,\AA) which we would not expect. It is more reasonable that this is a blend of both O VIII Ly-$\gamma$ and Fe XVII 3d-2p line emission, especially as we measure another Fe XVII 3d-2p line at 15.03\,\AA.

We have tentatively assigned the final two newly detected emission lines (found at 12.29 and 14.41\,\AA) to Fe XXI and Fe XVIII emission lines, although there are many possible alternative Fe transitions in these regions.

Of the emission lines we found, that were also measured by \cite{Guainazzi:2009fv}, intensity values in our 2004-07 stacked spectrum match well with \cite{Guainazzi:2009fv}'s values within 1$\sigma$ uncertainties, with a few exceptions: Ne IX r, Ne IX f, O VII RRC, Fe XVII 3s-2p and O VIII 1s-2p. For all the exceptions we measure higher intensities than \cite{Guainazzi:2009fv}. This is at least in part due to the difference in line widths assumed between the two works - \cite{Guainazzi:2009fv} assumed the lines were all unresolved (and therefore set their line widths to zero) whereas we allow the line widths to be free to vary.
For the Ne XI He-like triplet, which appears blended in our data, this leads our forbidden and resonance lines to broaden and account for more of the flux, and our intercombination line has a correspondingly smaller upper intensity limit than that of \cite{Guainazzi:2009fv} in response.
By fixing the widths of these lines to zero, our intensities fall much more in line with those of \cite{Guainazzi:2009fv}, and only the O VII RRC intensity is left with a discrepancy.

In the case of the O VII RRC, there is also potential contamination by the Fe XVII 2s-2p line (rest wavelength: 16.78\,\AA). This is discussed by \cite{Guainazzi:2009fv} who attribute an intensity of 0.9$\pm0.4 \times 10^{-5}$\,ph\,s$^{-1}$\,cm$^{-2}$ to the O VII RRC (given in their text) although their Table 2 quotes 0.47$\pm0.15 \times 10^{-5}$\,ph\,s$^{-1}$\,cm$^{-2}$ for the entire feature. An intensity of 0.9$\pm0.4 \times 10^{-5}$\,ph\,s$^{-1}$\,cm$^{-2}$ is consistent with our result for the entire feature at that position.

There are some emission lines which show hints of variation between our measured intensities in 2012-13 and 2004-07, but only at the 1$\sigma$ level. We return to comparing the two time periods in Sect. \ref{201213model_200407data_sect}.

The narrow emission lines have also been studied by \cite{2015MNRAS.453.2558N}, although their statistics is very low (partially due to extracting the spectrum from a smaller spatial area), especially over the O VII triplet region. Their best constrained lines are at higher energies (shorter wavelengths) than we investigate here.

\subsection{Feature at 23.2\,\AA}
\label{NCG1365_23A_feature}

In addition to the emission lines discussed in the previous section, we also detect a feature at $23.237 \pm 0.020$\,\AA\,in the 2012-13 stacked spectrum, which we initially identified as N VI 1s-5p (rest wavelength of $23.277$\,\AA), although only tentatively. 
The strength of this line is much higher than we would expect from a collisional or photoionised gas, because the lower order N VI transitions are much weaker than this feature (except the N VI He-like triplet).

Potentially, if there was charge exchange (CX) emission, this emission line could be enhanced compared to the lower order transitions. CX has been observed to produce X-rays within our Solar System, for example where the Solar Wind interacts with cometary comae, in Jupiter's aurorae and in the Earth's magnetosphere \citep{Dennerl:2010hm}, but recently evidence is building that this mechanism may be important in star formation regions as well \citep{2015RAA....15.2164L,Wang:2012he}.
If CX emission is enhancing the N VI 1s-5p line, then we would also expect the same to happen for ions of the same iso-electronic sequence in nearby elements, which is not the case; the O VII 1s-5p line (17.396\,\AA) and the Ne IX 1s-5p line (10.76\,\AA) are not enhanced.
Because of this, we consider the possibility of CX emission (or any other physical mechanism enhancing this transition) causing this feature to be very unlikely.

We note that the CCD on RGS2 covering this wavelength range failed shortly after launch, thus reducing the signal to noise ratio in the spectrum, and that there are known instrumental features from oxygen absorption \citep[23.05\,\AA\,and 23.35\,\AA;][]{2003A&A...404..959D} which also contribute in this region.

Finally, we estimate the significance of this feature to be 3$\sigma$, giving a probability of 0.3$\%$ that a feature of this strength may be seen by chance. This feature is driven by its central data point that, while strong, does have a larger uncertainty than the other data points around it. The above probability corresponds to an expectation that 1/370 events will be further than 3$\sigma$ from the correct model. Therefore given the number of bins we are using we would expect two non-physical features of this significance to be present in the spectrum.

Whilst we are not sure of the exact origin of this feature, we are not convinced that it represents a physical emission line from the source, and therefore do not consider it further within our analysis.

\subsection{Physically motivated emission models}
\label{physicalmodels_sect}

First we tested whether collisional ionisation models on their own could fit our spectra, as the starburst ring has previously been shown to contribute to the RGS spectra \citep{Guainazzi:2009fv}. We added collisional emission models one at a time (\cie models from {\sc spex}; over the best fit continuum and absorption model from Sect. \ref{gaussian_fit}), allowing their normalisations and temperatures to vary, until any further additions do not improve the overall fit. We considered the overall fit to be improved when the C-statistic decreased by $\geq 4.61$, corresponding to 90$\%$ confidence level for two free parameters (temperature and normalisation).
With this method we added four collisional phases, each giving a C-statistic decrease of 2975, 497, 32 and 40 respectively. These models correspond to best fit temperatures of 150$\pm$10\,eV 350$\pm$50\,eV, 660$\pm$30\,eV and 2800$_{-1000}^{+4100}$\,eV, similar to the temperature ranges of the two component collisional models in both \cite{Braito:2014ct} and \cite{Guainazzi:2009fv} (300\,eV-700\,eV and 150\,eV-1200\,eV, respectively) with an additional higher temperature phase. This high temperature phase is not well constrained and contributes very little to the line emission; only small contributions to O VIII 1s-2p and weak lines at the shortest wavelength range (11-12\,\AA) are made by this component. The total luminosity (0.3-2\,keV) from our four collisional models is 6.5$_{-1.7}^{+3.0}\,\times 10^{40}$\,erg\,s$^{-1}$, with a C-statistic of 1972.

\begin{figure}[!]
\centering
\resizebox{0.6\hsize}{!}{\includegraphics[angle=0]{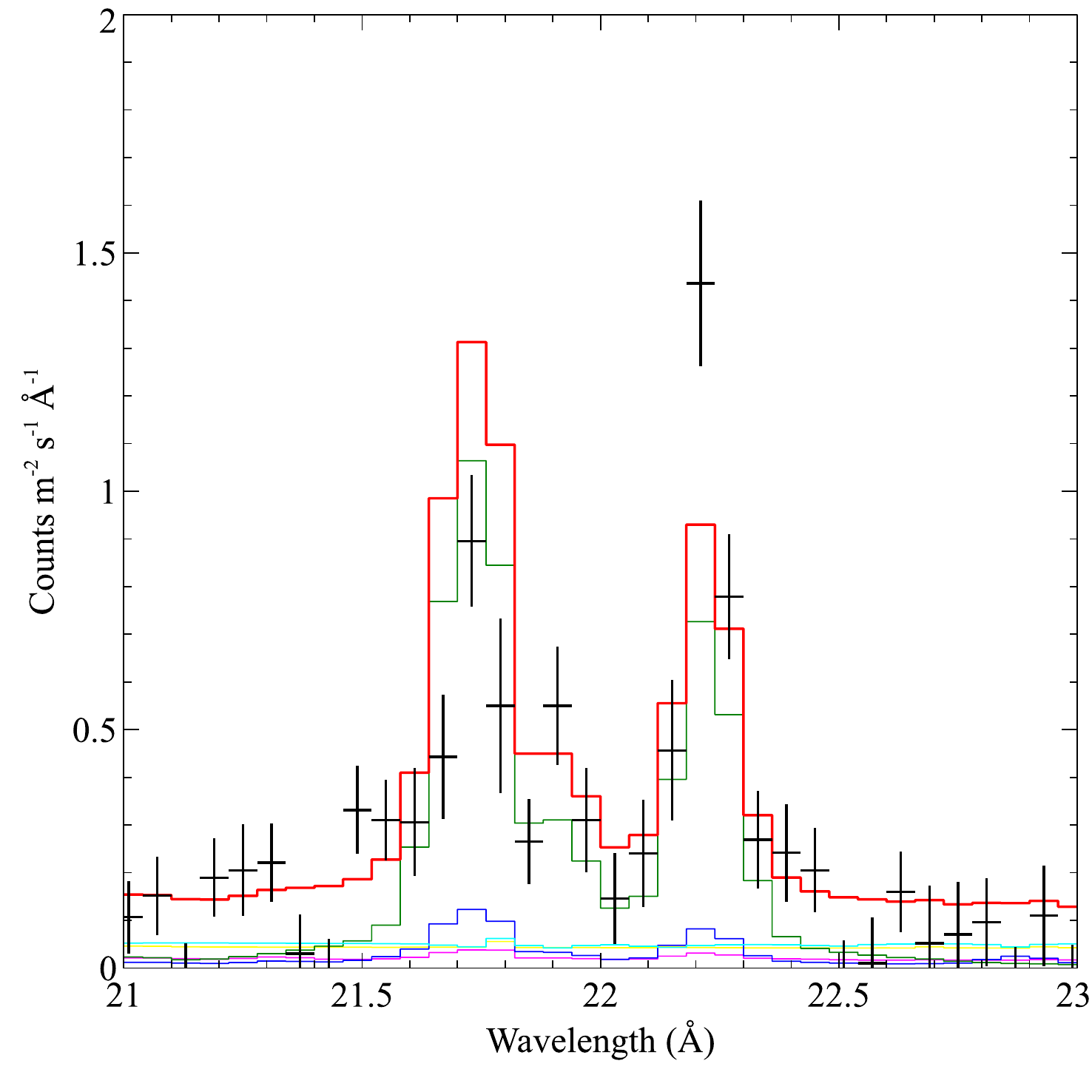}}
\caption{Best fit to the O VII triplet, with absorbed power law and four collisionally ionised emission models. The data are shown in black, the total model in red and the absorbed powerlaw component in light blue. The collisional emission components at temperatures 150\,eV, 350\,eV, 660\,eV and 2830\,eV are shown in green, dark blue, pink and yellow respectively. The data are binned by two for display purposes.}
\label{NGC1365_OVIIcieonly}
\end{figure}

Although no more collisional models are needed by the data, this is still not a satisfactory fit, especially around the O VII triplet ($\sim$22\,\AA), because the resonance line is overestimated and the forbidden line is underestimated (see Fig. \ref{NGC1365_OVIIcieonly}).

Alternatively we tried to fit the emission line spectra with only photoionised models.
As {\sc spex} does not yet include a photoionisation emission model, we fit the photoionised emission in our data with simulated spectra from {\sc Cloudy} \citep{2013RMxAA..49..137F}{.\footnote{After doing this work we discovered (through the Cloudy Yahoo Discussion Group; Topic 2624) that the ionisation parameter $\xi$ value used by {\sc Cloudy} 13.03 is defined over a different energy range (13.6\,eV to the high energy limit of the code) compared to the one used in the X-ray literature (13.6\,eV - 13.6\,keV). Using another form of the ionisation parameter ($U$) that {\sc Cloudy} uses we converted this into the `X-ray literature $\xi$' of each phase and found that the {\sc Cloudy} values are systematically 0.2 higher (in $\log\,\xi$) than the standard formulation used in X-ray literature (for the SED used in this work). This discrepancy between the two formulations of $\xi$ will be resolved in later versions of {\sc Cloudy}, as the next version of {\sc Cloudy} will be made to match the standard X-ray literature. The values currently quoted in this paper are the `{\sc Cloudy} $\xi$' values, for ease of comparison to any other work based on {\sc Cloudy} version 13.03 or previous.}}
To generate these models we assume a plane parallel geometry, with the central source illuminating the inner face of the cloud and with a flux density dependent on a chosen ionisation parameter, $\xi$. The gas cloud has proto-solar abundances \citep{2009LanB...4B...44L}. We vary the ionisation parameter ($\log\,\xi =$\,0.1-3.3\,erg\,cm\,s$^{-1}$), column density ($\log\,$N$_H =$\,21-24\,cm$^{-2}$) and turbulent velocity ($\log\,$v$_{turb} =$\,1.0-3.5\,\kms) of the illuminated cloud to generate a grid of simulated spectra. We use the reflected output of {\sc Cloudy} as our narrow emission line model.

We use a standard Seyfert 1 AGN SED
as we assume the X-ray narrow-line emitting clouds see the nuclear continuum.
This assumption is justified by the luminosity of the [O\,III] $\lambda$5007 emission line from the photoionised ionisation cone southeast of the nucleus; the photoionisation cone has a typical AGN [O\,III]\,/\,H$\alpha$ line ratio \citep{Veilleux:2003fe} and an [O\,III] $\lambda$5007 luminosity of $\sim 4 \times 10^{40}$\,erg\,s$^{-1}$ \citep{1997A&A...328..483K}, which together with the hard x-ray luminosity (14-195\,keV) of $10^{42.68}$\,erg\,s$^{-1}$ \citep{2010ApJS..186..378T}, place NGC 1365 within the expected region on a L$_{[OIII]}$\,(observed) vs L$_{14-195\,keV}$ plot \citep[see e.g.][]{2010ApJ...710..503W}. If the optical NLR was seeing an absorbed nuclear SED then the optical emission line luminosities would be much smaller than observed.

As these {\sc Cloudy} models only include the emission directly from the photoionised cloud we are modelling, we include in our fits the continuum and absorption described in Section \ref{gaussian_fit}. We fixed these components to their best fit parameters previously determined in Sect. \ref{gaussian_fit}.
Each {\sc Cloudy} model was imported (one by one) into {\sc spex} (as a \file model), and broadened by a \vgau model. The broadening is applied because while {\sc Cloudy} takes the effect of turbulence within the gas into account for the line strengths, it does not use this to broaden the line profiles in the produced {spectra \footnote{See Hazy 1 documentation, page 162}}. The normalisation of the \file model, and sigma value of the \vgau model are the only free parameters during this fit.
We fit each {\sc Cloudy} model in turn to the data using this process, and find a C-statistic value appropriate for each one.
We consider a first photoionisation model to be needed if the best fit C-statistic improves by 4.61, corresponding to a 90\,$\%$ confidence level for two extra free parameters (normalisation and broadening of the \file model). Further phases of photoionised gas are considered to be needed if the best fit C-statistic improves by 2.71, corresponding to a 90\,$\%$ confidence level for one additional free parameter (normalisation of the additional \file model; the broadening model is applied to all photoionised phases).

\begin{figure*}[!]
\centering
\resizebox{\hsize}{!}{\includegraphics[angle=0]{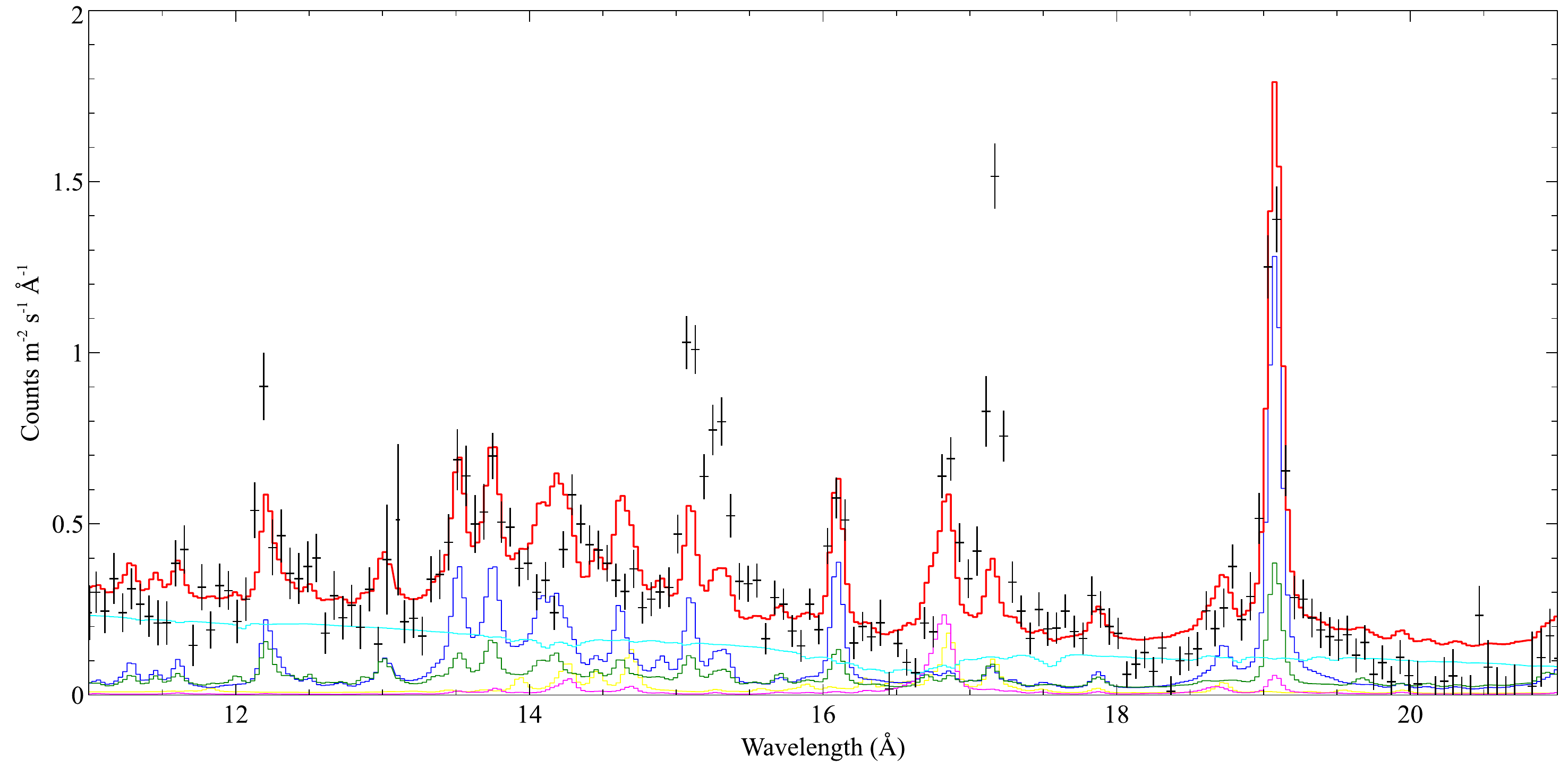}}
\caption{Best fit to the data with absorbed power law and four photoionised emission models, 11-21\,\AA\,range shown. The data are shown in black, the total model in red and the absorbed powerlaw component in light blue. The photoionised emission phases 1, 2, 3 and 4 are shown in green, pink, dark blue and yellow respectively.}
\label{NGC1365_OVIIphotoonly}
\end{figure*}

With this method we added four {\sc Cloudy} models, giving C-statistic decreases of 2161, 277, 116 and 52 respectively. These models found best fit $\log \xi$ values of 2.7, 1.1, 2.3 and 0.1, $\log N_H$ values of 23.5, 24, 21.25 and 24 cm$^{-2}$ and $\log v_{turb}$ values of 3.5, 1.0, 3.0 and 1.0 \kms respectively. The total luminosity (0.3-2\,keV) from our four {\sc Cloudy} models is 4.2$\pm0.6 \times 10^{40}$\,erg\,s$^{-1}$,  with a C-statistic of 2909. While no more photoionised models are needed by the data, this is still not a satisfactory fit, especially around the Fe lines at $\sim$15 and 17\,\AA\,(see Fig. \ref{NGC1365_OVIIphotoonly}).

As neither collisional nor photoionised emission alone fits the data, we then tested a combination of both. We add gas phases one by one;
we add either a collisional or photoionised phase at each stage, depending which gives a greater improvement in the C-statistic. We use the same criteria as above to determine whether another phase is necessary. The normalisations of all phases (both photoionised and collisional) are left free to vary throughout this process, along with the temperatures of the collisional phases and the broadening values.

\begin{figure*}[!]
\centering
\resizebox{\hsize}{!}{\includegraphics[angle=0]{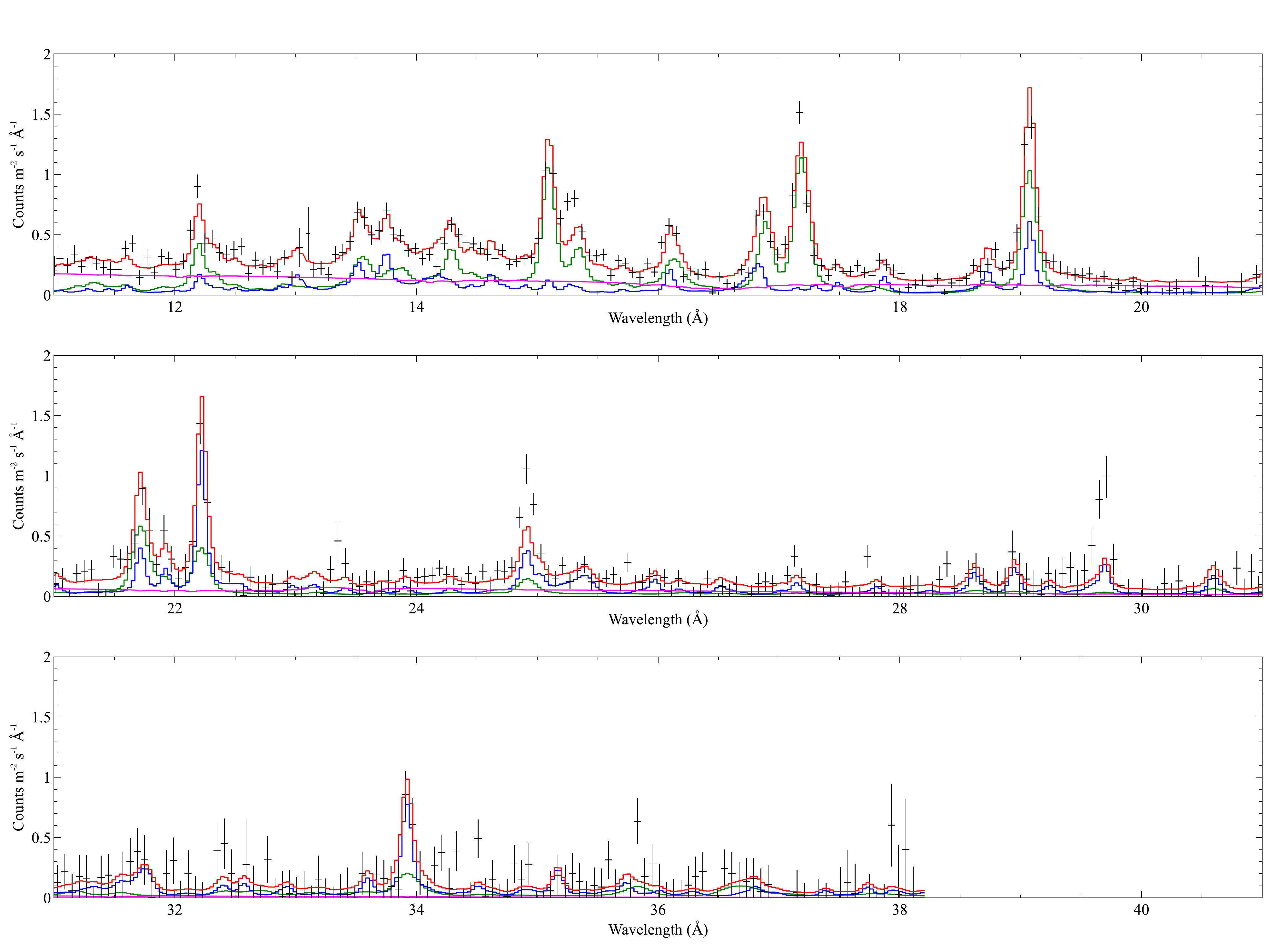}}
\caption{The stacked 2012-13 spectrum with best fit continuum and absorption model, two collisional emission models and three photoionised emission models, all using solar abundances. C-statistic $=$ 1756. The data are shown in black, the total model in red and the absorbed powerlaw component in pink. The collisional emission components are combined and shown in green. The photoionised emission phases are combined and shown in blue. The data are binned by two for display purposes. For details see text.}
\label{NGC1365_2C3Psolar}
\end{figure*}

The best fit in this case is obtained with two collisional phases (at temperatures 200$\pm$10 and 540$\pm$20\,eV) and three photoionised {\sc Cloudy} phases with $\log\,\xi$ of 1.5, 2.5 and 1.1 erg\,cm\,s$^{-1}$, $\log\,N_H$ of 22.25, 21.75 and 22.25 cm$^{-2}$, each having a $v_{turb}$ of $10^{2.5}$\,\kms, giving us a C-statistic of 1756. As you can see in Figure \ref{NGC1365_2C3Psolar}, the O VII triplet is now well fitted with this combination of two collisional and three photoionised emission components.
The total 0.1-2.5\,keV luminosity from the collisional components of this model is 2.45$\pm0.20 \times 10^{40}$\,erg\,s$^{-1}$, consistent with that expected from the starburst ring in this object, given by \cite{Wang:2009jh}'s \chandra analysis as 2.7$^{+0.7}_{-0.6} \times 10^{40}$\,erg\,s$^{-1}$.

Even though no further phases of either photoionised or collisionally ionised gas improve the statistical fit, and the fit is formally acceptable, the N VII Ly-$\alpha$ ($\sim25$\,\AA) and N VI forbidden ($\sim30$\,\AA) lines are still underpredicted by our combined model (Fig. \ref{NGC1365_2C3Psolar}). These are the only two strong nitrogen emission lines in this spectrum (although there are also other, much weaker nitrogen lines), which suggests there may be an overabundance of nitrogen compared to the proto-solar abundances \citep{2009LanB...4B...44L} used within our {\sc Cloudy} model.
This interpretation is supported by the detection of super-solar nitrogen abundances in other AGN outflows \citep[e.g.][using UV absorption lines]{2006ApJ...646..742G,2007ApJ...658..829A}, as well as a specific suggestion of higher nitrogen abundances in the optically detected outflows of this object by \cite{1999A&A...346..764S}.

To test this we ran further {\sc Cloudy} models, for the three phases we have already fitted, with a range of different nitrogen abundances ($\times$\,2, 5, 10 and 20 overabundant relative to the \cite{2009LanB...4B...44L} values we used previously). The varying nitrogen abundance alters the strength of the two nitrogen emission lines mentioned above: N VII Ly-$\alpha$ at $\sim$25 and N VI f at $\sim$30\,\AA. The two nitrogen RRCs also change slightly in strength, at $\sim$19 and $\sim$22.5\,\AA. The higher the nitrogen abundance, the stronger the nitrogen features, including the nitrogen RRC at $\sim$22.5\,\AA. As this particular feature gets stronger, the O VII f line gets weaker, which reduces the quality of the overall fit to the spectrum.

If there is an overabundance of nitrogen in the nuclear region of NGC 1365 we could reasonably expect it to be present in the starburst ring surrounding the nucleus as well as in the gas intrinsic to the AGN.

\begin{figure*}[!]
\centering
\resizebox{\hsize}{!}{\includegraphics[angle=0]{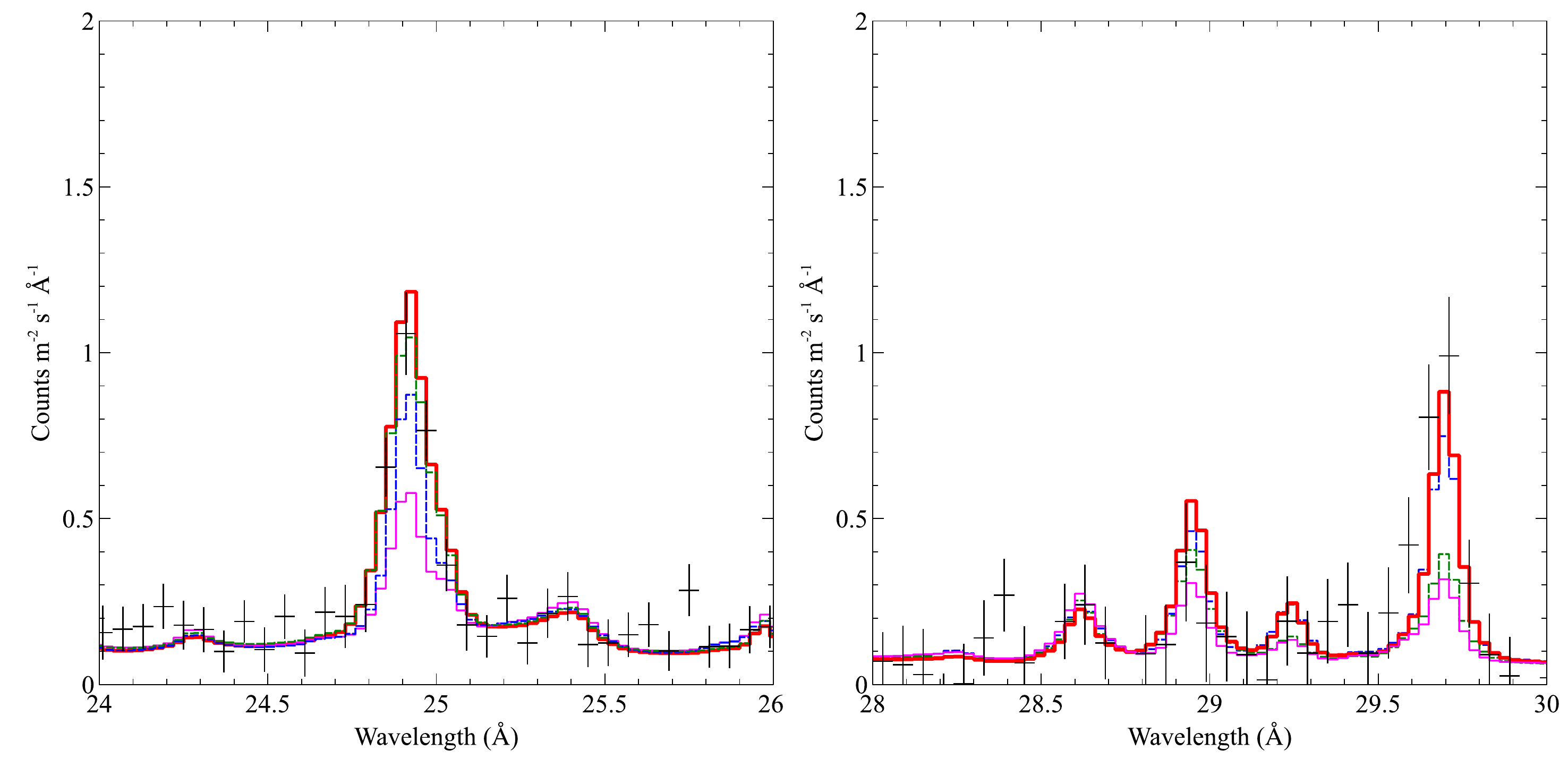}}
\caption{Fits with two collisional, three photoionised emission models. A five times over abundance of nitrogen model shown for intrinsic AGN gas only (dashed blue line), starburst emission lines only (dashed green line) and for both intrinsic AGN and starburst gas (thick red line), compared to the same model with solar abundances throughout (thin pink line). The data are binned by two for display purposes. For details see text and Table \ref{NGC1365_abundances_NTable}. Left: Region around N VII 1s-2p and N VI 1s-3p. Right: Region around N VI He-like triplet, including C VI 1s-3p.}
\label{NGC1365_N5_allcomps}
\end{figure*}

Therefore we tested 2, 5 and 10 $\times$ nitrogen abundances self-consistently with the collisional emission (representing the starburst region), the {\sc Cloudy} phases (representing the photoionised AGN emission) and the intrinsic AGN absorption (the traditional warm absorber phases) and found a nitrogen abundance of five times proto-solar fits the spectrum well, reducing the C-statistic by 120 to 1636 compared to the solar abundance model (see Table \ref{NGC1365_abundances_NTable} and Fig. \ref{NGC1365_N5_allcomps}).

\begin{table*}
\begin{minipage}[t]{\hsize}
\setlength{\extrarowheight}{3pt}
\caption{Results from testing different nitrogen abundances in both the AGN and starburst spectral components. Using solar abundances in all components gives C-statistic $=$ 1756. See text for details.}
\label{NGC1365_abundances_NTable}
\centering
\renewcommand{\footnoterule}{}
\begin{tabular}{l c c c}
\hline \hline
Components with & 2 $\times$ & 5 $\times$ & 10 $\times$ \\
non-solar abundances & nitrogen & nitrogen & nitrogen \\
\hline
intrinsic AGN emission & 1714 & 1674 & 1667 \\
\hline
intrinsic AGN emission & & & \\
and absorption & 1712 & 1663 & 1658 \\
\hline
starburst emission & 1713 & 1663 & 1701 \\
\hline
all & 1681 & 1636 & 1731 \\
\hline
\hline
\end{tabular}
\end{minipage}
\end{table*}

\begin{figure*}[!]
\centering
\resizebox{0.95\hsize}{!}{\includegraphics[angle=0]{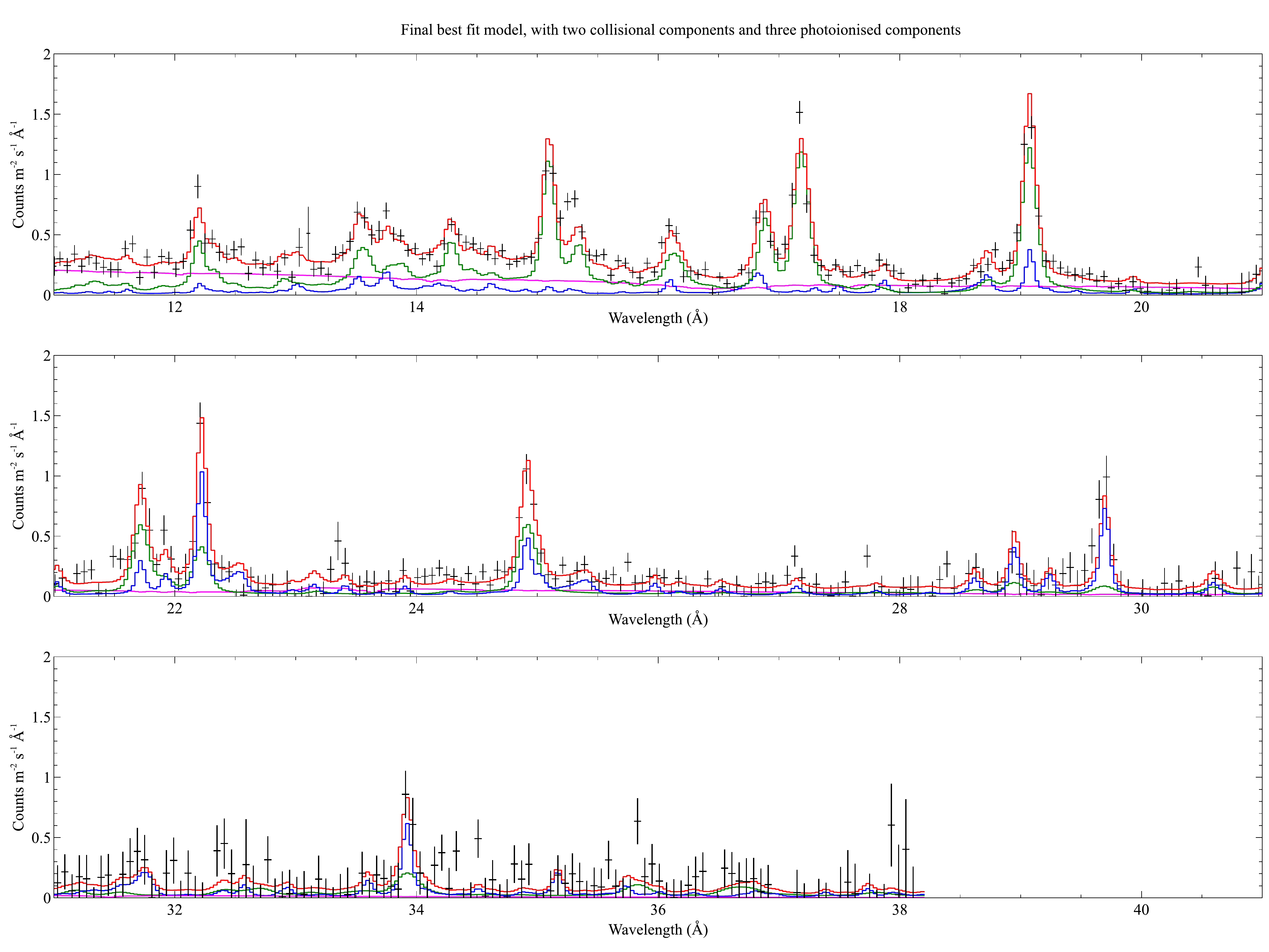}}
\caption{Final best-fit model for the 2012-13 stacked spectrum of NGC 1365 with 4.5\,$\times$\,solar nitrogen abundance, two collisional emission models representing the starburst ionised gas and three photoionised emission models representing the AGN ionised gas. The data are shown in black, the total model in red and the absorbed powerlaw component in pink. The collisional emission components are combined and shown in green. The photoionised emission phases are combined and shown in blue. The data are binned by two for display purposes. For details see text and Table \ref{NGC1365_finalpara_table}}
\label{NGC1365_N45_C2P3_all}
\end{figure*}

As you can see in Fig. \ref{NGC1365_N45_C2P3_all}, the nitrogen RRC at $\sim$22.5\,\AA\, visually seems stronger in this model than the data. This feature is from the photoionised emission phases, so one option to reduce the emission here in our model would be to reduce the nitrogen overabundance in the AGN emission. One could argue that leaving the AGN emission at solar abundances would help the model fit well here, but this `solution' would reduce the goodness of fit to the N VI triplet at $28.5-30$\,\AA, because the nitrogen overabundance in AGN (photoionised) emission contributes heavily to the N VI f emission line.

To narrow down the N abundance further, we tested (using the same method as above) 3, 4, 6, 7 and finally 4.5 $\times$ overabundance of nitrogen relative to the \cite{2009LanB...4B...44L} values. We conclude that a nitrogen abundance in the nuclear regions of NGC 1365 of 4.5$\pm$0.5 $\times$ solar \citep{2009LanB...4B...44L} best fits the data, giving a C-stat of 1634, and we show this model in Fig. \ref{NGC1365_N45_C2P3_all}. The collisional and photoionised phase parameters are contained in Table \ref{NGC1365_finalpara_table}.

As this method was successful at improving the fit of the nitrogen lines, we also tested higher iron abundances to improve the fit of the Fe XVII 3s-2p line at $\sim$17\,\AA\, and the Fe XVII 3d-2p / O VIII 1s-4p blend at $\sim$15.2\,\AA. Statistically, an overabundance of iron indicates a worse fit ($\Delta$C $>$ 34); the problematic iron lines remain underestimated and we now also have problems fitting lines from other elements. Therefore we do not include an overabundance of iron in our final model.

Our final model is shown in Fig. \ref{NGC1365_N45_C2P3_all} and includes two collisional emission phases, three photoionised emission phases, and an absorbed powerlaw continuum (as described in Sect. \ref{gaussian_fit}). All nuclear gas phases (both starburst and AGN, emission and absorption) have enhanced nitrogen abundances of 4.5 $\times$ solar.
The parameters of the emission phases in this model are shown in Table \ref{NGC1365_finalpara_table}.
The temperatures of our collisional phases are similar but not consistent with the collisional components from \cite{Guainazzi:2009fv}. This is due to different analysis methods, as we allowed our collisional phases' temperatures and normalisations to vary while fitting photoionised phases to the data, whereas \cite{Guainazzi:2009fv} kept their baseline model fixed.

\begin{table*}
\begin{minipage}[t]{\hsize}
\setlength{\extrarowheight}{3pt}
\caption{Parameters of the final best fit model, shown in Fig \ref{NGC1365_N45_C2P3_all}. C-statistic $=$ 1634. All components listed here have an enhanced nitrogen abundance of 4.5 $\times$ solar.}
\label{NGC1365_finalpara_table}
\centering
\renewcommand{\footnoterule}{}
\begin{tabular}{l c c c}
\hline \hline
Parameter & Phase 1 & Phase 2 & Phase 3 \\
\hline
Collisional gas & & & \\
T (keV) & 0.220$\pm0.010$ & 0.570$\pm0.015$ & - \\
\hline
Photoionised gas & & & \\
$\log\,\xi$ (erg\,cm\,s$^{-1}$)& 1.5$\pm0.2$ & 2.5$\pm0.2$ & 1.1$\pm0.2$ \\
$\log$\,N$_H$ (cm$^{-2}$) & 22.25$\pm0.25$ & 21.75$\pm0.25$ & 22.25$\pm0.25$ \\
v$_{turb}$ (\kms) & 10$^{2.5\pm0.5}$ & 10$^{2.5\pm0.5}$ & 10$^{2.5\pm0.5}$ \\
\hline
\hline
\end{tabular}
\end{minipage}
\end{table*}

We note that we find a better fit ($\Delta$C $=$ 120) when both the collisional line emission models are broadened by a {\sc spex} \vgau model. The collisional models fit best with a broadening of $\sigma=690\pm50$\,\kms, while the photoionised models are broadened by $\sigma <$225\,\kms. It is well known that RGS spectral lines can be broadened by the extent of a non point-like source. The amount of broadening is given by the formula:
\begin{equation}
\Delta\lambda = \frac{0.138}{m}\Delta\theta
\end{equation}
where $\Delta\lambda$ is the wavelength broadening (in \AA), $m$ is the spectral order and $\theta$ is the source extent in arcminutes \citep{SOC:2016vy}.

Using the \chandra images in \cite{Wang:2009jh} we estimate the soft X-ray emission of NGC 1365 to be 20\,arcseconds across, corresponding to a broadening of 620\,\kms at 22\,\AA. This is a reasonable match to the broadening of the collisional emission, and is much larger than that of the photoionised emission, indicating that most of the photoionised emission comes from a more compact region, within the starburst ring. By comparing RGS and \chandra ACIS spatially resolved spectroscopy, \cite{Guainazzi:2009fv} also concluded that the bulk of the photoionised emission comes from the central nucleus unresolved by \chandra.

\subsection{2012-13 best fit model with other data}
\label{201213model_200407data_sect}

\begin{figure*}[!]
\centering
\resizebox{\hsize}{!}{\includegraphics[angle=0]{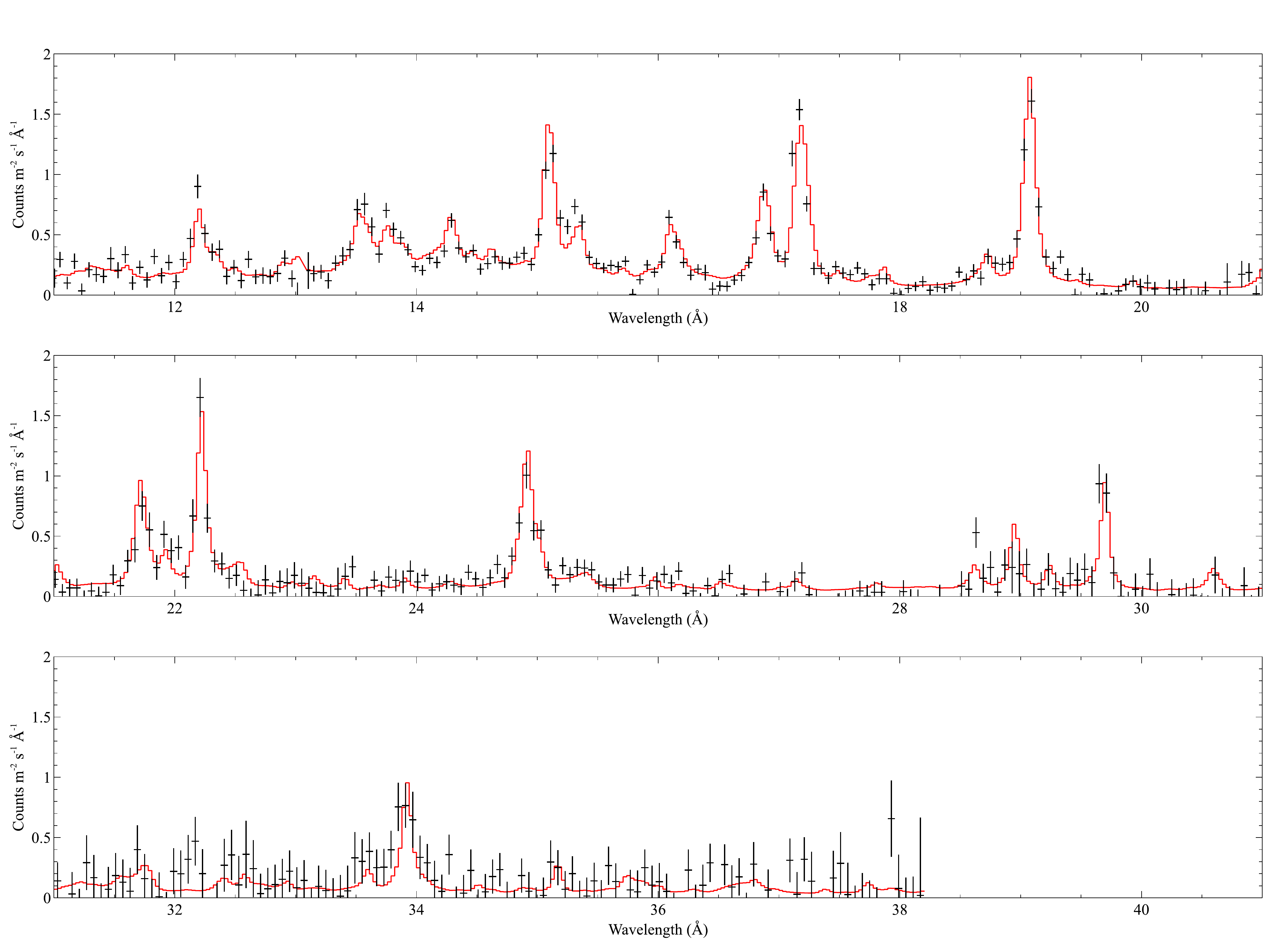}}
\caption{Our 2012-13 best-fit emission line model fitted to the 2004-07 stacked spectrum. Only the normalisations and broadening values of the emission line model were fitted to this spectrum. The data are shown in black and the total model in red. The data are binned by two for display purposes.}
\label{201213model_200407data}
\end{figure*}

We took the best fit model for our 2012-13 stacked spectrum, described above, and fit it to the 2004-07 stacked spectrum, with the 2004-07 continuum and absorption model from Sect \ref{gaussian_fit}. Only the normalisations and broadening values for our emission models were fitted. The temperatures of the collisional models, and the ionisation parameters, turbulent velocities and column densities of the photoionised models were all fixed to their best fit 2012-13 values. This gives an acceptable C-statistic of 1743, showing the emission is very similar between the two time periods (see Fig. \ref{201213model_200407data}). The broadening values fitted to the 2004-07 stacked spectrum are consistent with those found using the 2012-13 stacked spectrum (2004-07: 640$\pm$40\,\kms and $<$130\,\kms; 2012-13: 690$\pm$50\,\kms and $<$225\,\kms).
This confirms what is reported in Sect. \ref{gaussian_fit}, where we find no strong evidence for emission line changes between the two time periods when using gaussian models to fit the emission lines. 

We also did the same with the 2012-13 stacked spectral model and the third observation in that campaign (Jan 2013; not included in the 2012-13 stacked spectrum due to the differences in the intrinsic absorption compared to the other observations.). We used a continuum and absorption model fitted to this observation, by the same method discussed in Sect \ref{gaussian_fit}. The emisssion line model also fit successfully, with a C-statistic of 1461, confirming that the emission lines remained constant even as the absorption of the continuum changes. 

\section{Discussion}
\label{NGC1365_discussion}

We have modelled the narrow line emission from NGC 1365's nuclear region in a uniform and detailed way, using \xmm RGS data from 2004 to 2013. Firstly, using gaussian models for the emission lines we compared our results to the only other study which published individual emission line parameters in this wavelength range. We then fit the 2012-13 spectra with self-consistent, physical models, showing that a combination of both collisionally produced and photoionised emission lines reproduces the spectra well when the nitrogen abundance is increased by a factor of 4.5 relative to solar. This physical model also reproduces the earlier 2004-07 spectra without any temperature, ionisation parameter, turbulent velocity or column density changes in the emitting gas. 

We have not included photoexcitation followed by radiative decay in our models, which \cite{Kinkhabwala:2002fx} showed was important in enhancing higher order resonance lines in NGC 1068's soft X-ray emission line spectrum. \cite{Kinkhabwala:2002fx} also showed that the importance of photoexcitation decreases for higher ionic column densities, and at O VII radial column densities of $10^{19}$\,cm$^{-2}$ the spectrum imitates that of pure photoionisation followed by recombination/radiative cascade. The {\sc Cloudy} models we created show an O VII radial column density close to this $10^{19}$\,cm$^{-2}$ value, so we can ignore any photoexcitation contribution without affecting our results. 

In this section we discuss the relative contributions of our collisional and photoionised line emission and the enhanced 4.5 $\times$ solar abundance of nitrogen in the context of NGC 1365 as seen at other wavelengths.

\subsection{Starburst and AGN relative contributions}

The expected X-ray emission from star formation can be measured directly or estimated using measurements of infrared emission from the same region.
\cite{Wang:2009jh} measure L$_{0.3-10\,keV}$ luminosity as $1.87\pm0.36 \times 10^{40}$\,erg\,s$^{-1}$ from NGC 1365's nuclear starburst.
Using the total IR luminosity of NGC 1365's nuclear starburst measured by \cite{Tabatabaei:2013cc}, and the star formation IR to X-ray relation for galaxies in the local Universe from \cite{Symeonidis:2014cf}, we estimate a value consistent with this direct measurement from \chandra.

Collisional emission models at the temperatures we are considering here have very little emission at energies above the RGS band, so we can compare the \chandra 0.3-10\,keV luminosity to the luminosity we calculate from collisional emission models fitted to RGS data.

For the starburst (collisionally excited) emission in our final 2012-13 model we calculate a 0.3-10\,keV luminosity of 2.26$^{+0.17}_{-0.13}\,\times 10^{40}$\,erg\,s$^{-1}$, consistent with the \cite{Wang:2009jh} measurement.
For the 2004-07 data, we calculate a 0.3-10\,keV luminosity of 2.45$\pm0.11\,\times 10^{40}$\,erg\,s$^{-1}$, consistent with the \cite{Wang:2009jh} measurement at the 2$\sigma$ level

In the RGS band we have analysed (11-38\,\AA), 
the photoionised emission contributes 25-46$\%$ of the total narrow line emission, with collisional emission contributing the rest. In addition, the collisional components contribute a very small amount of emission to the continuum, which is not included in the estimate above.
In fact the photoionised emission is 5-15$\%$ of the total soft X-ray emission in this range. This estimate is consistent with that of \cite{Guainazzi:2009fv} ($\leq 10 \%$) with the possibility of a slightly higher contribution. We attribute this to a slight difference in analysis method; \cite{Guainazzi:2009fv} fix their collisional emission model first, and then estimate that the photoionised emission contributes any `leftover' line intensities, whereas in this work we allow the collisional emission phases to vary in both normalisation and temperature while fitting photoionised phases (see Sect. \ref{physicalmodels_sect}).

\subsection{Higher nitrogen abundances}

Abundance measurements for AGN are usually made using broad emission lines in rest-frame UV \citep[e.g.][]{2002ApJ...564..592H}, narrow absorption lines in UV \citep[absolute abundances can be measured because hydrogen is observed, e.g.][]{2007ApJ...658..829A}, and narrow absorption lines in X-ray \cite[where only relative abundances of metals can be measured, e.g.][]{Steenbrugge:2011kg}.
We are aware of few previous studies estimating AGN elemental abundances using X-ray narrow emission lines. \cite{2004mas..conf..191J} looked at three objects (NGC 1068, NGC 4051 and NGC 4507) and determined that NGC 1068 (Seyfert 2) has a $\leq 2$ times overabundance of nitrogen, NGC 4051 (Narrow Line Seyfert 1) has an overabundance of neon (but the factor is undetermined) and NGC 4507 (Seyfert 2) is consistent with solar abundances.
Both \cite{Kinkhabwala:2002fx} and \cite{2002A&A...396..761B} have also found evidence for 2-3 $\times$ solar abundance of nitrogen in the emission line spectrum of NGC 1068 (using RGS and \chandra LETGS data, respectively).
RGS observations of starburst galaxies have been used to estimate metal abundances from the soft X-ray collisional emission lines. \cite{2002MNRAS.335L..36R} find supersolar abundances of magnesium, silicon, neon and nitrogen (between 2 and 5 $\times$ solar), and \cite{2004ApJ...606..862O} also find supersolar abundances of magnesium, silicon and sulphur, although to a lesser extent (up to 1.5 $\times$ solar). \cite{2004ApJ...606..862O} only analyse the RGS data up to 20\,\AA, so they do not measure any nitrogen abundance.

As well as analysing the images, \cite{Wang:2009jh} also model the collisional soft X-ray emission using \chandra ACIS spectra and present abundances. They find a large spatial variation in abundance values, with super solar neon, magnesium, silicon and iron abundances in the starburst ring and very subsolar oxygen, neon, magnesium, silicon and iron abundances in the diffuse regions around the starburst. In both these regions the abundances were determined using only a one temperature collisional emission model (the ring model also included a power law component), and using low resolution ACIS spectra (compared to the \chandra gratings and \xmm RGS).
It has been shown that a one temperature collisional model at sub solar abundances and a two temperature collisional model at (or near) solar abundances cannot be distinguished at CCD resolution \citep{1994ApJ...427...86B,1998MNRAS.296..977B}, and as the \cite{Wang:2009jh} study took place before the publication of any high resolution soft X-ray spectra of this object, we suspect this may be the case here.
Without the resolution to determine the strengths of and ratios between individual emission lines, distinguishing the contribution of collisional and photoionised emission to the soft X-ray spectrum is subject to great uncertainties, and therefore we consider these abundances also to be uncertain.
As a check, we used the collisional model parameters given in \cite{Wang:2009jh}'s Table 2 for their ``diffuse'' and ``ring$+$'' regions and found these models reproduce the 11-21\,\AA\,region of our 2012-13 RGS data well, but do not reproduce the emission longwards of 21\,\AA, including the O VII and N VI He-like triplets and the N VII and C VI Ly-$\alpha$ emission lines.

As we find a best fit spectral model with 4.5$\pm$0.5 times enhanced abundance of nitrogen in the nuclear area of NGC 1365, it is natural to ask why and where this might come from.

One possibility is that the enhanced strength of nitrogen lines may be caused by the Bowen fluorescence mechanism, as described by \cite{Sako:2003ft}, rather than a higher nitrogen abundance.
\cite{Sako:2003ft} shows that at high optical depths, resonance line photons (e.g. O VIII Ly-$\alpha$ doublet at 18.967 and 18.973\,\AA) can be absorbed by other ions, exciting an electron to a higher energy level, when the receiving transition has a very similar energy (e.g. N VII Ly-$\zeta$ at 18.974\,\AA). This electron then undergoes radiative cascade and therefore releases photons associated with lower order transitions of the receiving ion. 
The overall effect of this is to reduce the apparent abundance of the initial ion (in our example case, O VIII) and increase the apparent abundance of the receiving ion (in our example case, N VII).
While this mechanism can enhance the strength of lower order N VII emission lines, it does not do the same for N VI emission lines, which are enhanced in NGC 1365's spectrum. Therefore Bowen fluorescence is not a satisfactory explanation for our observations in this case.

The alternative explanation for the enhanced nitrogen lines we observe is that they are caused by an over abundance of nitrogen in the source. If the nitrogen abundance in NGC 1365's nucleus is enhanced, then signatures of this should be visible in data at other wavelengths too. As NGC 1365 is an obscured AGN, the UV emission is too weak for the usual types of abundance studies.
However, using optical \hst data, \cite{1999A&A...346..764S} modelled a photoionised region of optical narrow-line emission and found evidence for a factor of three increase in nitrogen abundance.
Their major problem with this conclusion was that they thought the 2-10\,keV luminosity (L$_{2-10\,keV}$) of the AGN was too low for the number of ionising photons needed for their model, and considered the necessary absorption column of $\sim$10$^{23}$\,cm$^{-2}$ to reconcile this to be too high.
With the extended energy range and higher resolution spectra available from  both \chandra and \xmm, we now know that this level of absorption is not unusual for NGC 1365, avoiding this problem with the \cite{1999A&A...346..764S} optical narrow-line emission model, and therefore supporting the need for a higher than solar nitrogen abundance.

Most of the nitrogen in the ISM today is ejected during the AGB phase of low and intermediate mass stars (1-8\,M$_{\odot}$), with negligible oxygen enrichment. During the red giant branch and asymptotic giant branch phases of intermediate mass stars, dredge up episodes can occur. The outer convective envelope reaches down to inner regions where nitrogen has been produced by the CNO cycle. This, combined with `hot-bottom burning', which converts some or most of the carbon at the base of the convective zone to nitrogen, enhances the surface abundance of nitrogen. The nitrogen is then ejected into the ISM through winds driving mass loss from these stars. Unfortunately, stellar yields of nitrogen are still reasonably uncertain, as they are sensitive to convection models and estimates of dredge up efficiency \citep{2009nceg.book.....P}. 

For this mechanism to be the cause of our observed higher nitrogen abundance, the star formation history of NGC 1365 would have to greatly deviate from average over a long period of time. Without any significantly unusual other features, this seems to be an unlikely explanation, especially as the N/O ratio in outer H II regions of this galaxy is measured to be consistent with solar \citep{Bresolin:2005jl}.

The other option for increasing the nitrogen abundance is a shorter timescale, smaller scale effect. The only nucleosynthetic environment in the nucleus is the starburst region, so if the nitrogen enhancement is created in the nucleus, then the starburst region may well be the cause.
The presence of Wolf-Rayet (WR) stars has been shown to correlate with a higher N/O abundance ratio by \cite{Brinchmann:2008jq}, using an SDSS sample with 570 WR galaxies and more than 1000 WR galaxy candidates. WR stars are thought to descend from O stars, after losing their outer envelope \citep{2000eaa..book.....M}.
WR stars were detected by \cite{1992ApJ...395L..91P} in spectra of a `hot spot' near the nucleus of NGC 1365, so this could be enhancing the nitrogen abundance in our measurements.

\subsection{Long term behaviour and distance of emitting gas}

Comparing the 2004-07 and 2012-13 spectra shows no evidence of variability of the emission lines. Each individual emission line is consistent at a 3$\sigma$ uncertainty level between the two epochs (Table \ref{NLgaussian_table}).
The individual starburst and AGN emission components also have consistent overall luminosities (11-38\,\AA) between the two epochs.
This lack of variability means we are unable to infer the size of the emitting region or its distance from the ionising continuum.

We can also calculate an estimate of the location of the bulk of the photoionised gas by using the broadening of the photoionised phases. As discussed in Sect. \ref{physicalmodels_sect}, we find an upper limit of $<225$\,\kms to the broadening of the photoionised gas using the 2012-13 spectrum. The 2004-07 spectrum gives us a consistent, and tighter, constraint of $<130$\,\kms.

Assuming that any broadening is due to bulk motion of the gas orbiting the central source, we can calculate a lower limit to the distance. Using a black hole mass of $2 \times 10^{6} M_{\odot}$ \citep{Risaliti:2009cx}, the upper limits on the line widths imply that the emitting gas must be located further than 0.5\,pc from the central source.

This is consistent with the estimate by \cite{Guainazzi:2009fv}, using their best-fit ionisation parameter, that the emitting gas is located $\geq 0.75$\,pc from the central source.
Our distance is different from that of \cite{Braito:2014ct}, who conlcude that emission line gas is located at $\sim 10^{15}$\,cm ($3 \times 10^{-4}$\,pc) from the central source, calculated from tentative variability of Mg XII emission lines. In fact, if we extend our model down to the Mg XII Ly-$\alpha$ line at 8.42\,\AA, we find that our collisional emission models contribute most of the intensity of this line.
This much closer distance estimate is also based on shorter term variability (within an observation) than we investigate in this work, and as discussed above, we do not find evidence of variability over these longer timescales.

We can obtain an upper limit to the size of the photoionised emission line region from its angular extent on the sky. From the RGS, which is a slitless spectrograph, spatially extended emission causes broadening of the emission lines, as discussed in Sect. \ref{physicalmodels_sect}.
Using the 2004-07 constraint we calculate a photoionised gas extent of $<$\,4.2", and at the distance of NGC 1365 this converts to a physical projected distance of $<$\,380\,pc.
This is consistent with \chandra results from \cite{2015MNRAS.453.2558N}, who show that extracting spectra from a circular aperture of 10" diameter centred on the nucleus increases the proportion of photoionised emission and correspondingly reduces the collisional emission contribution.
In addition, \cite{Wang:2009jh} fit the starburst ring and surrounding diffuse emission with collisional emission models while photoionised emission is expected to be the dominant component in the inner region around the nucleus, within a radius of 10". 

We are more likely to be observing NGC 1365's nucleus side on, rather than face on (using the unification scheme and NGC 1365's classification range of Seyfert 1.5-2). Therefore we can assume the 380\,pc represents the maximum extent of two ionisation cones in opposite directions from the nucleus, and we can deproject one half of that, to estimate the physical distance of the bulk of the emitting gas from the central source. Using a range of viewing angles from 80\degree-\,45\degree\,(if 0\degree\, represents looking face on), we calculate a range of upper limits for the distance of 190-270\,pc between the central source and the photoionised gas.

This upper limit is entirely consistent with previous distance estimates for the photoionised gas (discussed in Sect. \ref{NGC1365_intro}), and is the first upper limit estimate for the distance of X-ray emitting photoionised gas within this object.

\section{Conclusion}
\label{NGC1365_conclusion}

This is the most detailed spectral description of the emission lines from NGC 1365's nucleus so far, enabled by the quantity of good quality data available in the \xmm archive.

In this work we show that the X-ray narrow emission line spectrum of NGC 1365 is well represented by a combination of two collisionally ionised (kT of 220$\pm10$ and 570$\pm15$\,eV) and three photoionised ($\log \xi$ of 1.5$\pm0.2$, 2.5$\pm0.2$, 1.1$\pm0.2$) phases of emitting gas, all with higher than solar nitrogen abundances. We attribute the collisionally ionised gas to the starburst surrounding the nuclear region and the photoionised gas to the Seyfert 2 nucleus in the centre.

This physical model is the best fit to the 2012-13 stacked spectrum, and yet also fits well to the 2004-07 stacked spectrum, without changing any characteristics of the emitting gas phases.
Our finding of 4.5$\pm$0.5 times solar nitrogen abundance in the nuclear region of this system represents the second time an over abundance of nitrogen has been suggested in this source \citep[the first being][from optical measurements]{1999A&A...346..764S}.

We find that the photoionised X-ray emitting gas is $0.5 \leq $ distance $ \leq 270$\,pc from the ionising continuum of the central source.

\begin{acknowledgements}
The authors thank J. Kaastra for help with the \xmm RGS data reduction process used and I. Ferreras for useful discussions. This work is based on observations obtained with \xmm, an ESA science mission with instruments and contributions directly funded by ESA Member States and the USA (NASA). M. W. acknowledges the support of a PhD studentship awarded by the UK Science \& Technology Facilities Council (STFC).

\end{acknowledgements}

\bibliography{ngc1365biblio}

\end{document}